\documentclass[apj]{emulateapj}

\usepackage{graphicx}
\usepackage{multirow}
\usepackage{epsfig}
\usepackage{amsmath}
\slugcomment{Submitted to The ApJL}
\shorttitle{The Flattening of Dust Attenuation Curve to $z=2.5$}
\shortauthors{Li et al.}

\newcommand\sersic{S$\acute{\rm e}$rsic }

\begin{document}

\title{The Flattening of Dust Attenuation Curve to $z=2.5$}

\author{Yubin~Li\altaffilmark{1,2}, Xian~Zhong~Zheng\altaffilmark{1}, Feng~Shan~Liu\altaffilmark{3}}

\altaffiltext{1}{Purple Mountain Observatory, Chinese Academy of Sciences, 2 West Beijing Road, Nanjing 210008, China; xzzheng@pmo.ac.cn}
\altaffiltext{2}{University of Chinese Academy of Sciences, 19A Yuquan Road, Beijing 100049, China}
\altaffiltext{3}{College of Physical Science and Technology, Shenyang Normal University, Shenyang 110034, China}

\begin{abstract}

We  examine the evolution of dust attenuation curve using a sample of 9504 disk star-forming galaxies (SFGs) selected from the CANDELS and 3D-HST surveys and a new technique relying on the fact that disk SFGs of similar stellar masses at the same cosmic epoch are statistically identical in stellar populations.  We attribute the discrepancy in median magnitude between face-on ($b/a>0.7$) and edge-on ($b/a\leq0.4$) subsamples solely to dust attenuation, and obtain the average attenuation in the rest-frame UV and optical as functions of stellar mass and redshift out to $z=2.5$.  Our results show that the attenuation curve becomes remarkably flatter at increasing redshift for both massive and low-mass disk SFGs, and remains likely unchanged with galaxy stellar mass at a fixed epoch  within uncertainties. Compared with the Calzetti law, our dust attenuation curves appear to be slightly steeper at $0.5<z<1.4$ and remarkably flatter at $1.4<z<2.5$. 
Our findings are consistent with a picture in which the evolution of dust grain size distribution is mainly responsible for the evolution of the dust attenuation curve in SFGs; dust shattering becomes  a dominant process at $z<\sim 1.4$, resulting in an enrichment of small dust grains and consequently a steeper attenuation curve. We stress that extinction correction for high-$z$ galaxies should be done using mass- and redshift-dependent attenuation curves. 

\end{abstract}

\section{INTRODUCTION} \label{sec:intro}

Dust attenuation curve is a key observable to understanding of dust evolution and metal cycle, and to the observability of stellar populations in galaxies. The attenuation curve is governed by abundance, grain size distribution and chemical composition of dust, which plays a critical role in multiple physical processes regulating the evolution of galaxies, including gas cooling, condense of molecular clouds, star formation, chemical enrichment, as well as radiation transfer of the ultraviolet (UV) and optical into the middle- and far-infrared emission in galaxies. While the attenuation law for the local starburst galaxies \citep{Calzetti00} has been widely used in studies of nearby and distant galaxies,  a systematic investigation of dust attenuation curves in high-$z$ star-forming galaxies (SFGs) is still missing. 

The production of dust from core-collapse type\,II supernovae  and asymptotic giant branch (AGB) stars is correlated with star formation history of galaxies \citep[e.g.,][]{Valiante09,McKinnon16}. In contrast, star formation in the interstellar medium (ISM) substantially reduces dust content and locks metals in stars.  The size distribution of dust grains in ISM is continuously shaped by grain growth processes via accretion of gas-phase metals or adhesion of dust grains, and also by destruction mechanisms through heating by stellar winds or ionizing radiation \citep[e.g.,][]{Nozawa13,Asano13}.  Dust shattering is predicted to control the size distribution of dust grains at a timescale of $t\sim \tau_{\rm SF}^{1/2}$\,Gyr after the peak of star formation, and accordingly leaving the dust attenuation curve steeper \citep{Asano14}.  Measurements of dust attenuation curves across cosmic time enable us to draw essential constraints on dust evolution \citep{Santini14,Bekki15}. 

Efforts have been made to determine dust attenuation curves in high-$z$ SFGs \citep{Buat12,Kriek13,Reddy15,Zeimann15,Salmon16} and quasars \citep[e.g.][]{Gallerani10}. The measures are derived from the UV-selected galaxies \citep{Buat12}, emission-line galaxies \citep{Zeimann15} or a handful of galaxies with spectroscopic observations spreading over wide stellar mass and redshift ranges \citep{Kriek13,Reddy15}, and thus unlikely representative for either the global population  or a mass-complete subpopulation of SFGs. The evolution of dust attenuation curve out to high-$z$ is still controversial,  partially due to the selection bias and the correlations with spectral type and specific star formation rate (sSFR) \citep[e.g.,][]{Kriek13}. Moreover, the  attenuation measurements yet are mostly made through SED modeling, and apparently affected by the degeneracies between attenuation, stellar age and metallicity.

\begin{figure*}
 \begin{center}
  \includegraphics[width=0.8\textwidth]{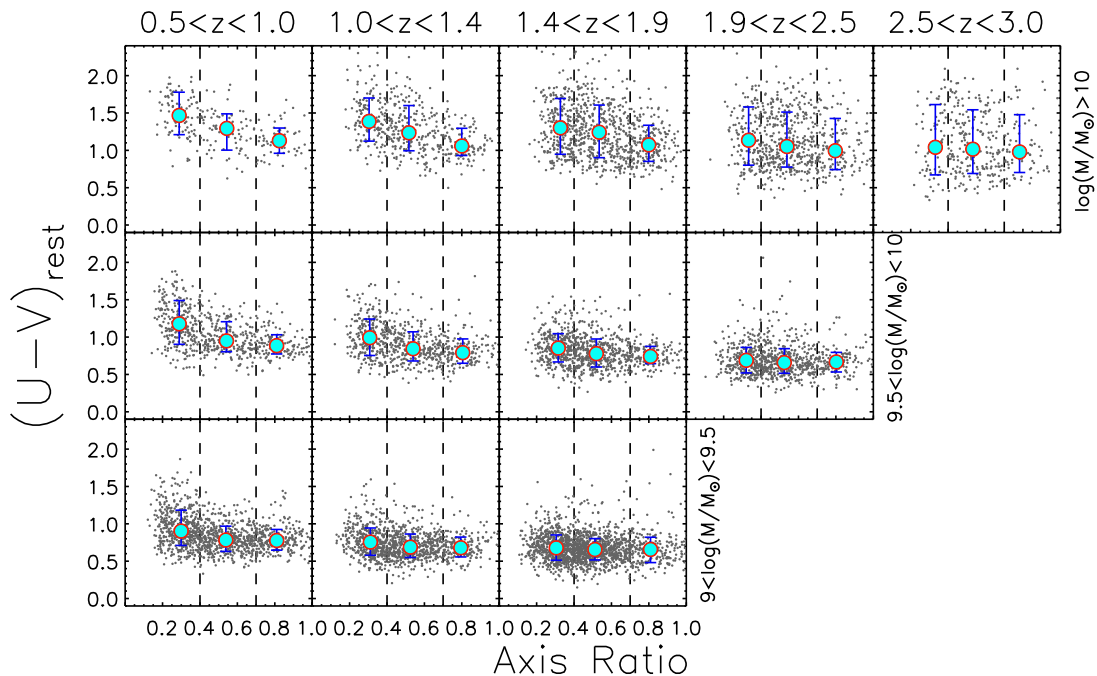}
  \caption{Distribution of rest-frame $U-V$ color as a function of axis ratio for our sample of  9504 disk SFGs.  In each bin, median $U-V$ is derived for subsamples of edge-on ($b/a\leq0.4$), intermediately-inclined ($0.4< b/a\leq 0.7$) and face-on ($b/a$>0.7), as shown by the large cyan circles. The errorbars indicate the 16th and 84th percentiles of the $U-V$  distribution. } 
  \label{fig:f1}
 \end{center}
\end{figure*}

Instead, we develop a new technique to derive the average attenuation curve of a subpopulation of SFGs at the same cosmic epoch in a statistical manner. We apply this technique to the CANDELS+3D-HST data sets, and draw the dust attenuation curves for mass-selected disk SFGs over $0.5<z<3$. 
The paper is organized as follows.
The data and sample selection are described in Section~\ref{sec:data}. The details of our technique used to derive the attenuation curve and our results are presented in Section~\ref{sec:results}. We discuss and summarize our results in Section~\ref{sec:disc}. 
Throughout this work, we adopt a cosmology of $H_0=70$\,km\,s$^{-1}$\,Mpc$^{-1}$, $\Omega_{\rm M}=0.3$ and $\Omega_\Lambda=0.7$.

\begin{table}
 \begin{center} 
  \caption{Number of sample galaxies \label{tab:sample}}
  \begin{tabular}{cccc}
   \hline\hline
   \multirow{2}{*}{$z$}& \multicolumn{3}{c}{$\log (M_\ast/M_\odot)$} \\ 
   \cline{2-4}
     & $9.0 - 9.5$ & $9.5 - 10$ & $>$10  \\
   \hline
     $0.5 - 1.0$ & 1213  & 659 & 236 \\
     $1.0 - 1.4$ & 1080  & 684 & 478 \\
     $1.4 - 1.9$ & 1641  & 826 & 818 \\
     $1.9 - 2.5$ & ...     & 836 & 677 \\
     $2.5 - 3.0$ &  ... & ... & 356 \\
\hline\hline
\end{tabular}
 \end{center}
\end{table}

\section{Data and Sample Selection} \label{sec:data}

This study makes use of public data in the GOODS-N, GOODS-S, COSMOS, AEGIS and UDS fields covered by the observations from CANDELS \citep{Grogin11, Koekemoer11} and 3D-HST \citep{Brammer12}.  The multi-band catalogs including redshift, rest-frame photometry, and stellar masses have been provided by the 3D-HST team \citep[see][for details]{Skelton14,Momcheva15}. The prior redshift is chosen in the order of spectroscopic , grism and photometric redshift. The rest-frame photometry of each galaxy is derived at the best redshift using the EAZY software tool \citep{Brammer08}. 
The stellar masses are estimated from  the multi-band photometric catalogs using FAST \citep{Kriek09}.  
Galaxy structural parameters, including the \sersic index ($n$), effective radius ($r_{\rm e}$) and axis ratio ($b/a$) are from \cite{VanderWel14}, securely extracted from \textit{HST}/WFC3 F160W images with GALFIT \citep{Peng02}.

We aim to select  disk galaxies of similar properties to derive dust attenuation curves. Firstly, the $UVJ$ selection \citep{Whitaker11} is adopted to separate star-forming galaxies (SFGs) from quiescent galaxies over the redshift range $0.5<z<3.0$.   Secondly, we identify the extended SFGs with good-fit \sersic index $n<1.5$ as disk SFGs,  for which axis ratio is a direct probe of inclination angle.  We also limit the disk SFGs of given stellar mass and redshift to be comparable in size (effective radius along semi-major axis). This is done via narrowing down the effective radius within 1\,$\sigma$ centered at the median ($r_{\rm e,0}$) of the log-normal distribution of size at the given stellar mass.  This cut excludes those SFGs that are either compact ($\log r_{\rm e}<\log r_{\rm e,0}-1\,\sigma$) or diffuse ($\log r_{\rm e}>\log r_{\rm e,0}+1\,\sigma$), differing from the extended disk SFGs in many aspects.  The size limitation is more important for high-$z$ where many SFGs appear to be compact and  systematically redder and rounder than the normal disk SFGs.
We also add stellar mass cuts  $>10^9\, M_{\odot}$ at $z<1.9$, $>10^{9.5}\, M_{\odot}$ at  $1.9<z<2.5$ and $>10^{10}\, M_{\odot}$ at $2.5<z<3.0$ for blue SFGs to ensure the selection completeness in stellar mass \citep[see also][]{VanderWel14}. The mass completeness limits roughly correspond to F160W\,$\sim$ 24.5\,mag for blue galaxies, which is a conservative magnitude limit to derive a robust GALFIT measurements.  
Finally, we select a sample of 9504 disk SFGs in the redshift range $0.5<z<3.0$. The sample galaxies are split into 5 redshift bins and 3 stellar mass bins to ensure that each bin has sufficient objects for statistics.   Table~\ref{tab:sample} lists the numbers of objects in these bins. 

Our selection avoids the bias in magnitude selection against dusty galaxies.  
Although the three bins close to the cut edge contains 14\%$-$40\% of objects with  $F160W>24.5$\,mag, only  $<6\%$ are  $F160W>25.5$\,mag. Their structural parameters estimated using GALFIT still have an accuracy of about 20\% for disks with $24.5<F160W<25.5$\citep{VanderWel12,VanderWel14}. We caveat that heavily-obscured disk SFGs may be too faint to be missed by our selection. We argue that such galaxies are rare and do not cause a significant underestimate of dust attenuation because they marginally alter the median magnitude.

\begin{figure*}
 \begin{center}
  \includegraphics[width=0.8\textwidth]{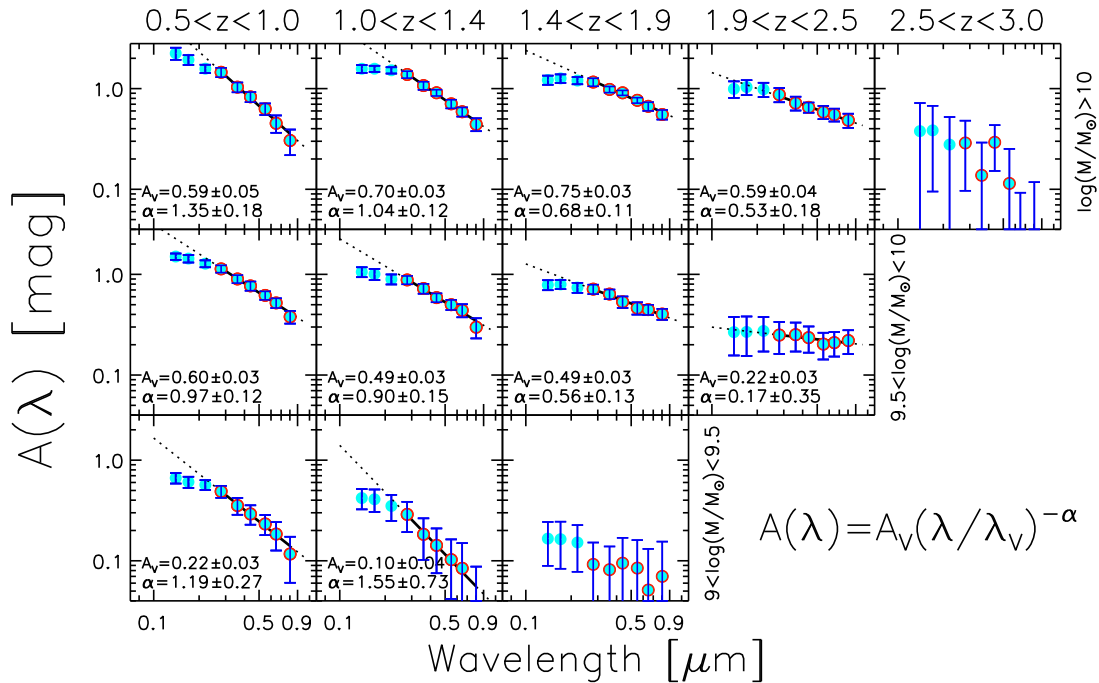}
  \caption{Dust attenuation curves in disk SFGs divided into 3 stellar mass bins and 5 redshift bins.}
  \label{fig:f2}
 \end{center}
\end{figure*}

\section{ANALYSIS AND RESULTS} \label{sec:results}

Our sample disk SFGs are selected by morphology ($n<1.5$ and $r_{\rm e} \sim r_{\rm e,0}\pm1\,\sigma$) and color.  The sample galaxies in a given bin are of similar stellar masses and redshifts can be treat as the same population. Their  axis ratio $b/a$ is a direct measure of the inclination angle of these disks  projected on the sky. Accounting for the thickness of disks with $n<1.5$, we refer the disks with $b/a\leq0.4$ to be edge-on and those with $b/a>0.7$ to be face-on.  Distant disk SFGs of similar stellar masses and redshifts are nearly identical in sSFR (i.e. stellar populations) \citep{Rodighiero11}.  
The two subsamples are statistically characterized by the same attenuation curve. 
 The edge-on  and face-on disk SFGs are statistically indistinguishable from each other in terms of their intrinsic properties, including spectral energy distribution (SED), metallicity and composition of interstellar medium. The difference in geometry  between edge-on and face-on causes discrepancy in opacity, and consequently shaping the observed SEDs by different amount of dust attenuation. 
The luminosity discrepancy between the two subpopulations as a function of wavelength thus describes the dust attenuation curve in the disk SFGs of given stellar mass and redshift.

The rest-frame ultraviolet (UV) and optical, including 1400\,\AA, 1700\,\AA, 2200\,\AA, 2800\,\AA, $U$, $B$, $V$, $R$ and $I$-band, absolute magnitudes (i.e. luminosities) from the 3D-HST+CANDELS catalogs are used to examine the luminosity discrepancy between edge-on and face-on subpopulations.   Since dust attenuation usually makes the UV-to-optical SED of a galaxy to become dimmer and redder,  We thus examine the correlations between color and inclination angle. 
Figure~\ref{fig:f1} shows the distribution of rest-frame $U-V$ as functions of axis ratio, stellar mass and redshift for our sample of 9504 disk SFGs. The median values of $U-V$ are presented for face-on , intermediately-inclined ($0.4< b/a\leq 0.7$) and edge-on subsamples in each of bins divided by stellar mass and redshift. The errorbars represent the 16th and 84th percentiles of the $U-V$  spread of a  subsample. It is clear that in each bin $U-V$ systematically increases at decreasing $b/a$ from face-on to edge-on.  The stellar mass bins of $9<\log(M_\ast/M_\odot)<9.5$ at $z>1.9$ and of $9.5<\log(M_\ast/M_\odot)<10$ at $z>2.5$ are not included because of the incompleteness issue. 
The $U-V$ discrepancy between edge-on and face-on decreases at decreasing stellar mass and at increasing redshift. 
The results from Figure~\ref{fig:f1} confirm that edge-on disk SFGs are systematically redder than face-on ones at at $z<2.5$, consistent with the expected effects by dust attenuation.

Comparing the median absolute magnitudes  in the rest-frame UV and optical between edge-on and face-on subpopulations, we derive magnitude discrepancy in these bands and plot them in Figure~\ref{fig:f2}.  We estimate uncertainties in magnitude discrepancy using bootstrapping method. The uncertainties are globally decided by the magnitude spreads and object numbers of edge-on and face-on subsamples.  We point out that for subsamples with $9<\log(M_\ast/M_\odot)<9.5$ the data points exhibit large errorbars due to the small magnitude difference between the two subpopulations although the object numbers are large in these subsamples. As we mentioned above, the magnitude discrepancy can be solely attributed to dust attenuation of galaxies viewed at edge-on relative to at face-on in a population-averaged sense. We accordingly refer the magnitude discrepancy between edge-on and face-on subsamples to be attenuation magnitude $A(\lambda)$. The data points in Figure~\ref{fig:f2} thus give the dust attenuation curves in disk SFGs divided by stellar mass and redshift. We note that edge-on and face-on subpopulations have the same attenuation law and $A(\lambda)$ in Figure~\ref{fig:f2} quantifies the attenuation difference between the two subpopulations, following the same form of the attenuation curve. 
It is clearly seen that $A(\lambda)$ decreases with wavelength in all panels.  

To quantitatively describe these attenuation curves, we fit the data points with a power-law function 
\begin{equation}
A(\lambda)=A_{\rm V}\left(\frac{\lambda}{\lambda_V}\right)^{-\alpha},
\end{equation}
where $\lambda_V=0.55\micron$ and $A_{\rm V}$ is the attenuation magnitude in the $V$-band.  The power index $\alpha$ measures the steepness of the attenuation curve.  We realize that the actual extinction law is more complex than a single power-law, as shown by the formula from \cite{Calzetti00} for local starburst galaxies. Our method to derive the extinction curve is based on comparison between edge-on and face-on disk SFGs in a statistical manner, and lacks the resolution to determine high-order variation beyond the power-law \cite[see also][]{Shao07,Chevallard13}.  We only fit the data points from the 2800\,\AA\ to $I$-band. These rest-frame bands are covered by deep imaging  in the observed optical and near-infrared window (0.4$-$5\,$\micron$) over the redshift range $0.5<z<3$ examined here. And the rest-frame far-UV bands (1400, 1700 and 2200\,\AA) are significantly dimmed by dust attenuation and photometry in the corresponding observed bands is often contaminated by noise due to  the low signal-to-noise ratio. We thus exclude the three far-UV bands from our fitting because they tend to be strongly biased.  

The best-fit $A_{\rm V}$ and $\alpha$ are derived using the least squares method in logarithm space, and the results are shown in Figure~\ref{fig:f2}. We can see that the steepness described by the power index $\alpha$ decreases with increasing redshift, particularly at $z>1.4$. This is to say that the dust attenuation curve becomes flattening  at higher redshift. The normalization of the attenuation curve $A_{\rm V}$ remains nearly unchanged with redshift for $\log(M_\ast/M_\odot)>10$ but decreases for $\log(M_\ast/M_\odot)<10$. At a fixed redshift, the steepness appears to be similar within uncertainties, while the normalization declines for lower-mass disk SFGs.

\begin{table}
 \begin{center} 
  \caption{The deviation  of the dust attenuation curve from the Calzetti law \label{tab:fit2}}
   \begin{tabular}{cccc}
   \hline\hline
   \multirow{2}{*}{$z$}& \multicolumn{3}{c}{$\log (M_\ast/M_\odot)$} \\ 
   \cline{2-4}
     & $9.0 - 9.5$ & $9.5 - 10$ & $>$10  \\
   \hline
     $0.5 - 1.0$ & $-0.32\pm0.26$ & $0.08\pm0.12$ & $-0.30\pm0.17$ \\
     $1.0 - 1.4$ & $-0.42\pm0.47$ & $0.02\pm0.17$ & $-0.14\pm0.11$  \\
     $1.4 - 1.9$ & ...   & $0.44\pm0.13$ & $0.33\pm0.09$ \\
     $1.9 - 2.5$ & ...   & $0.82\pm0.38$ & $0.35\pm0.19$   \\
\hline\hline
\end{tabular}
 \end{center}
\end{table}

\begin{figure}
 \begin{center}
  \includegraphics[width=0.45\textwidth]{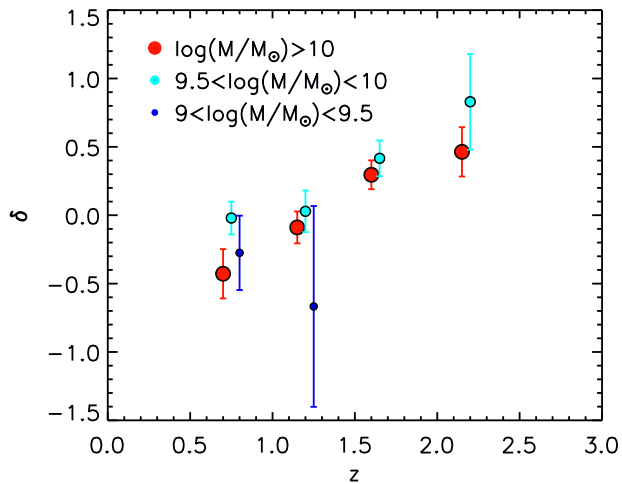}
  \caption{The slope deviation of dust attenuation curve from that of the Calzetti law as functions of stellar mass and redshift.  }
  \label{fig:f3}
 \end{center}
\end{figure}

The dust attenuation curves in high-$z$ galaxies were often compared with the Calzetti law to see the evolutionary effects\citep{Buat12,Kriek13,Zeimann15}.  Similarly, we also quantify the deviation of our attenuation curves from the Calzetti law following \cite{Noll09}: 
\begin{equation}
A(\lambda)=\frac{A_{\rm V}}{4.05} k^{'}(\lambda) (\frac{\lambda}{\lambda_V})^\delta, 
\end{equation}
where $\delta$ describes the deviation of attenuation curve $A(\lambda)$ from the Calzetti law $k^{'}(\lambda)$. One will obtain $\delta>0$ if the attenuation curve is flatter than the Calzetti law, and $\delta<0$ if the attenuation curve is steeper.   We ignore the extinction bump at 2175\,\AA\ originally included in the formula of \cite{Noll09}.  

We perform fitting in the logarithm space using the linear least squares method and obtain the best-fit $\delta$ for 10 attenuation curves, as shown in Table~\ref{tab:fit2}.  Figure~\ref{fig:f3} demonstrates the deviations as a function of redshift. We can see that the attenuation curve is remarkably flatter than the Calzetti law before $z\sim1.4$, and the attenuation curve is slightly steeper than the Calzetti law  at $z<\sim1.4$.

\section{DISCUSSION AND SUMMARY} \label{sec:disc}

We develop a new technique to derive dust attenuation curves in distant disk SFGs.  Statistically speaking, the carefully-selected disk SFGs  appear to have identical stellar populations (i.e. sSFR) and thus have similar intrinsic luminosities in the UV and optical bands. Due to the disk geometric structure, the observed luminosity of a disk galaxy is  more attenuated at increasing inclination angle. The magnitude discrepancy between face-on and edge-on can be attributed to dust attenuation, determined by the opacity that depends on gas column density, gas-to-dust ratio,  dust grain compositions, and size distribution.   These parameters are regulated by different physical processes and may vary with redshift and galaxy stellar mass.  It is challenging to resolve contributions from these  processes.  This is the basis of our technique and can be measured in a statistical manner.  
Our technique avoids the degeneracies between dust attenuation, stellar age and metallicity, and the assumption of dust-star geometry included in  the methods based on SED modeling. 
We select  a sample of 9504 extended disk SFGs with  $n<1.5$ and $\log (M/M_\odot)>9$ over $0.5<z<3$ from the CANDELS and 3D-HST surveys.   
The sample is divided into three mass bins and five redshift bins. By comparing the median magnitude between edge-on ($b/a\leq 0.4$) and face-on  ($b/a>0.7$) subsamples in each bin, we obtain the average attenuation in the rest-frame 2800\,\AA, $U, B, V, R$ and $I$-band as functions of stellar mass and redshift out to $z=2.5$ and derive the average attenuation curve.

It has been established that at a fixed stellar mass, SFGs appear to have higher sSFR at increasing redshift out to $z\sim 2.5$; and at a fixed cosmic epoch, disk-like SFGs obey a constant sSFR \citep{Whitaker15}.  On the other hand,  gas fraction is also expected to increase with redshift in terms of the Kennicutt-Schmidt Law.  We find that attenuation $A_{\rm V}$ declines with galaxy stellar mass at all epochs from $z=0.5$ to $z=2.5$ as shown in Figure~\ref{fig:f2}.  This is not surprising because less massive SFGs tend to be lower in metallicity and less dusty, leading to a smaller opacity discrepancy between face-on and edge-on, and smaller attenuation. 
Moreover, the attenuation $A_{\rm V}$ evolves little for massive disk SFGs with $\log(M_\ast/M_\odot)>10$, but decreases with redshift for low-mass ones with $\log(M_\ast/M_\odot)<10$. Taking $A_{\rm V}$ as an indicator of opacity, the little evolution of $A_{\rm V}$ with decreasing redshift suggests that multiple processes, including the decline of gas fraction and the chemical enrichment in ISM (alternatively, the decrease of gas-to-dust ratio), balance each other and leave the dust content roughly constant in the massive disk SFGs.  In contrast, the low-mass disk SFGs follow the same decline of gas fraction, but need a more rapid evolution in metallicity in order to yield an increase of opacity (dust content) with decreasing redshift.  This is reasonable because massive SFGs are able to get self-enriched within a short timescale even at early cosmic epochs but the chemical enrichment in low-mass SFGs is inefficient due to their shallow potential well and largely influenced by the enriched intergalactic medium in the low-$z$ universe.  \cite{Whitaker14} pointed out that the infrared excess ($L_{\rm IR}/L_{\rm UV}$) of low-mass SFGs with $\log(M_\ast/M_\odot)<10.5$ remains nearly constant  out to $z=2.5$ but increases rapidly with redshift for massive SFGs, suggestive of a relative higher dust content in massive SFGs at high-$z$. This is consistent with our results.

Part of SFGs at high-$z$ are found to be elongated, differing from a disk structure \citep{VanderWel14b}.   
This would potentially bias the attenuation discrepancy between the edge-on and face-on subsamples to be smaller.  We argue that such galaxies tend to be round with higher \sersic indices. our morphological criteria to select SFGs of similar disk structures and to much extent get rid of such elongated SFGs, supported by the similarity of distributions in axis ratio.

The most striking result is that the attenuation curve becomes remarkably flatter at increasing redshift out to $z=2.5$ for both massive and low-mass disk SFGs, and remains unchanged  within uncertainties with galaxy stellar mass at fixed epochs.  The quantitative comparison in Figure~\ref{fig:f3} shows that the attenuation curve in disk SFGs is flatter than the Calzetti law at $1.4<z<2.5$, and is roughly identical to or steeper than the Calzetti law at $0.5<z<1.4$.  Since the determination of attenuation curves in the wavelength range from 2800\,\AA\ to the $I$-band is insensitive to dust compositions (e.g., 2175\AA\ bump),  the steepness of the curves is mostly decided by dust grain size distribution. Our finding thus indicates that the dust grains become systematically smaller from $z=2.5$ to $z=0.5$. This is consistent with the model prediction by \cite{Asano14}, in which dust is dominated by large grains produced by SNe\,II and AGB stars at high redshifts; shattering powered by ionizing photons and collisions controls the dust evolution in later time, resulting in a gradually higher fraction of small grains in the total;  and coagulation effects may eventually become dominant when the shattering gets weakened along with the evolved stellar populations.  If so,  dust grains would grow bigger and the attenuation curve would become steeper. Indeed,  the Calzetti Law, i.e. the dust attenuation curve in local starburst galaxies,  is slightly flatter than the derived curves at $0.5<z<1.4$, as shown in Figure~\ref{fig:f3}.

Compared with previous works, our technique is totally independent on the degeneracies between stellar age, metallicity and attenuation, and escapes from the assumptions of intrinsic stellar populations and their SEDs \citep{Buat12,Kriek13,Zeimann15} and uncertainties in estimate of SFR \citep{Penner15}.  Still, previous results on the evolution of dust attenuation curve can be better understood based on our findings.  For instance, The anti-correlation between the steepness of dust attenuation curve and sSFR among  SFGs at $0.5<z<2.0$ from \cite{Kriek13} is actually caused by the rapid evolution in both sSFR and steepness over the redshift range examined.   \cite{Zeimann15} found that $z\sim2$ emission line galaxies have a shallower attenuation curve than that of local starburst galaxies,  in agreement with our results. 
\cite{Buat12} presented that 20\% of their UV-selected sample galaxies at $0.95<z<2$ have a dust attenuation curve steeper than the Calzetti law and this fraction increases to 40\% for the infrared-detected galaxies.  In terms of our finding that the steepness of dust attenuation curve evolves with redshift from steeper to flatter relative to the Calzetti law over $0.5<z<2.5$, the change of fraction is purely due to selection effect because the infrared observations are limited to detect more galaxies at $z<1.4$.

In short, we present a first attempt to decompose dust evolution with cosmic time and its dependence on galaxy stellar mass.  This is critical to understanding physical processes driving dust evolution.  Our conclusions are based on  mass-complete subsamples and thus representative for the populations of high-$z$ SFGs.  We  stress that extinction correction for high-$z$ galaxies should be done using a matched attenuation curve in terms of galaxy stellar mass and cosmic epoch.

\acknowledgments

This work is based on observations taken by the 3D-HST Treasury Program (GO 12177 and 12328) with the NASA/ESA HST, which is operated by the Association of Universities for Research in Astronomy, Inc., under NASA contract NAS5-26555. 
This work is supported by the Strategic Priority Research Program ``The Emergence of Cosmological Structures'' of the Chinese Academy of Sciences (grant No. XDB09000000), the National Basic Research Program of China (973 Program 2013CB834900) and National Natural Science Foundation of China through grant U1331110 and 11573017.

\end{document}